\documentclass[twocolumn,preprintnumbers,amsmath,amssymb,showpacs,prb,aps]{revtex4}
\usepackage{txfonts}
\usepackage{pifont}
\usepackage{amssymb}
\usepackage{dcolumn}
\usepackage{amsmath}
\usepackage[dvips]{epsfig}

\makeatletter
\def\btt#1{\texttt{\@backslashchar#1}}%
\DeclareRobustCommand\bblash{\btt{\@backslashchar}}%
\makeatother

\begin{document}
\title{Competition of crystal field splitting and Hund's rule coupling in
two-orbital magnetic metal-insulator transitions}
\author{Ya-Min Quan$^{1}$, Liang-Jian Zou$^{1,2 \footnote{Correspondence author,
        Electronic mail: zou@theory.issp.ac.cn}}$ Dayong Liu$^{1}$,and Hai-Qing Lin$^{2}$}
\affiliation{ \it $^1$ Key Laboratory of Materials Physics,
              Institute of Solid State Physics, Chinese Academy of Sciences,
               P. O. Box 1129, Hefei 230031, China\\
         \it $^2$ Department of Physics, Chinese University of Hong Kong,
                  Shatin, New Territory, Hong Kong, China\\}
\date{Nov 4, 2011}

\begin{abstract}
Competition of crystal field splitting and Hund's rule coupling in
magnetic metal-insulator transitions of half-filled two-orbital
Hubbard model is investigated by multi-orbital slave-boson mean
field theory. We show that with the increase of Coulomb correlation,
the system firstly transits from a paramagnetic (PM) metal to a {\it
N\'{e}el} antiferromagnetic (AFM) Mott insulator, or a nonmagnetic
orbital insulator, depending on the competition of crystal field
splitting and the Hund's rule coupling. The different AFM Mott
insulator, PM metal and orbital insulating phase are none, partially
and fully orbital polarized, respectively. For a small $J_{H}$ and a
finite crystal field, the orbital insulator is robust. Although the
system is nonmagnetic, the phase boundary of the orbital insulator
transition obviously shifts to the small $U$ regime after the
magnetic correlations is taken into account. These results
demonstrate that large crystal field splitting favors the formation
of the orbital insulating phase, while large Hund's rule coupling
tends to destroy it, driving the low-spin to high-spin transition.
\end{abstract}

\pacs{71.30.+h, 75.30.Kz, 71.10.Hf, 71.27.+a, 71.10.Fd}
\maketitle

\section{INTRODUCTION}

Mott-Hubbard metal-insulator transitions (MIT) in multi-orbital
Hubbard models have been extensively studied since there are rich
phase diagrams \cite{Liebsch, Bouadim2009, Akihisa2004, Philipp2007,
Eberhard2009, Medici2005, Song2009, Medici2009, Ruegg2005,
DaiXi2006, Qimiao2010, Anisimov2002}. A few of many-body  approaches
have been employed to study the paramagnetic Mott transition, such
as dynamic mean field theory (DMFT)\cite{Liebsch, Bouadim2009,
Akihisa2004, Philipp2007, Eberhard2009, Medici2005, Song2009,
Medici2009} and Kotliar-Ruckenstein slave-boson method
\cite{Ruegg2005, DaiXi2006}, \emph{etc}. With the slave spin
technique, Yu and Si recently showed that a paramagnetic orbital
insulating phase can exist when the Hund's coupling approaches zero
\cite{Qimiao2010}. However, generally speaking, MIT transitions are
usually accompanied by magnetic transitions. How spin correlation
affects the nature of MIT is seldom discussed.

On the other hand, the crystal field splitting plays an important
role in the properties of multi-orbital systems. It is not clear
whether the interplay between crystal field splitting and spin
exchange splitting is a key factor for antiferromagnetic (AFM) MIT.
Though Hasegawa \cite{Hasegawa1997} studied the spin correlation
effect in half-filled doubly degenerate Hubbard model and found the
Hund's coupling plays a very considerable role in magnetization, he
did not find the MIT at half filling. It deserves to explore the
influence of crystal field splitting on the MIT in the presence of
magnetism. And how the crystal field and spin correlation modify the
orbital insulating phase is also an interesting issue.

In this paper, the multi-orbital Kotliar-Ruckenstein slave-boson
method is generalized with spin degree of freedom. The effects of
crystal field splitting and Hund's rule coupling on Mott transition
are also discussed. We find that besides  $J_{H}=0$, orbital
insulator can exist for small enough but finite  $J_{H}$, different
from Yu and Si's results \cite{Qimiao2010}. The magnetic phase
diagram as a function of $U$ and crystal field splitting shows that
the phase boundary can be seriously changed by the Hund's rule
coupling and crystal field splitting, demonstrating the contrary
role of these two factors. The rest of this paper is organized as
follows: in Sec.II we describe the model Hamiltonian and theoretical
approach. The effects of crystal field splitting are discussed and
corresponding phase diagram are presented in Sec. III. The final
section is devoted to the conclusion remarks.

\section{Model Hamiltonian and Methods}
The two-orbital Hubbard model adopted in this paper is as follows:
\begin{eqnarray}
\label{eq:Hamiltonian}
 H&=&H_{0}+H_{I}\\
   H_{0}&=&-\sum_{ij\alpha\beta\sigma}\left(t_{\alpha\beta}c^{\dagger}_{i\alpha
   \sigma}c_{j\beta\sigma}+h.c.\right)+ \sum_{i\alpha\sigma}\left(\varepsilon_{i
   \alpha\sigma}-\mu\right)n_{i\alpha\sigma}  \\
H_{i}&=&U\sum_{i\alpha}n_{i\alpha\uparrow}n_{i\alpha\downarrow}+\sum^{\left(
  \alpha>\beta\right)}_{i\sigma\sigma^{\prime}}
   \left(U^{\prime}-J_{H}\delta_{\sigma\sigma^{\prime}}\right)n_{i\alpha\sigma}
   n_{i\beta\sigma^{\prime}}               \nonumber\\
   & &-J_{H}\sum_{i\alpha\neq\beta}\left(c^{\dagger}_{i\alpha\uparrow}c_{i\alpha
   \downarrow}c^{\dagger}_{i\beta\downarrow}c_{i\beta\uparrow}
   -c^{\dagger}_{i\alpha\uparrow}c^{\dagger}_{i\alpha\downarrow}c_{i\beta
   \downarrow}c_{i\beta\uparrow}\right),
\end{eqnarray}
where $c^{\dag}_{i\alpha\sigma}$ is a creation operator of an
electron with the orbital index $\alpha$ and spin $\sigma$  at the
lattice site $i$ , $n_{i\alpha\sigma}$ is the corresponding
occupation number operator.  The hopping integral between orbitals
$\alpha$  and $\beta$ is denoted by $t_{\alpha\beta}$. The intraband
(inter-band) Coulomb repulsion and Hund's rule coupling are denoted
by $U$  ($U^{\prime}$ ) and $J_{H}$, respectively. Here we set
$U^{\prime}=U-2J_{H}$. We introduce new boson operators $e, p, d, b,
t$ and $q$, which denote the empty, single occupation, double
occupation in two orbital, double occupation in one orbital, triple
occupation, and quadruple occupation, respectively. The physical
electron creation operator $c^{\dag}_{i\alpha\sigma}$ is represented
as $Z_{i\alpha\sigma}f^{\dag}_{i\alpha\sigma}$, where
$f^{\dag}_{i\alpha\sigma}$ is quasi-particle creation operator. The
renormalization factor $Z_{i\alpha\sigma}$ is given by
\begin{eqnarray}
\label{eq:renormalization}
  z_{i\alpha\sigma} &=& \hat{Q}_{i\alpha\sigma}^{\frac{-1}{2}}
  \left(p^{\dag}_{i\alpha\sigma}e_{i}+
  b^{\dag}_{i\alpha}p_{i\alpha\bar{\sigma}}
  +\sum_{\sigma'}d^{\dag}_{i\sigma_{\alpha}\sigma'_{\beta}}p_{i\beta\sigma'}
  +t^{\dag}_{i\alpha\sigma}b_{i\beta}\right. \nonumber\\
  & &+ \left.\sum_{\sigma'}
  t^{\dag}_{i\beta\sigma'}d_{i\bar{\sigma}_{\alpha}\sigma'_{\beta}}
  +q^{\dag}_{i}t_{i\alpha\bar{\sigma}}\right)
  (1-\hat{Q}_{i\alpha\sigma})^{\frac{-1}{2}},
\end{eqnarray}
where
\begin{eqnarray}
\label{eq:particle number}
\hat{Q}_{i\alpha\sigma}&=&p^{\dag}_{i\alpha\sigma}p_{i\alpha\sigma}
+b^{\dag}_{i\alpha}b_{i\alpha}+\sum_{\sigma^{\prime}}d^{\dag}_{i\sigma_{
\alpha}\sigma^{\prime}_{\beta}}d_{i\sigma_{\alpha}\sigma^{\prime}_{\beta}}+
\sum_{\sigma^{\prime}}t^{\dag}_{i\beta\sigma^{\prime}}t_{i\beta\sigma^{\prime}}
\nonumber\\
& &+t^{\dag}_{i\alpha\sigma}t_{i\alpha\sigma} +q^{\dag}_{i}q_{i}.
\end{eqnarray}

In the present slave boson states, the normalization constraint and
Fermi number constraint maintaining physical Hilbert space are as
follows:
\begin{eqnarray}
\label{eq:normalization}
1&=&e_{i}^{\dag}e_{i}+\sum_{\alpha\sigma}(p^{\dag}_{i\alpha\sigma}p_{i\alpha\sigma}
+t^{\dag}_{i\alpha\sigma}t_{i\alpha\sigma})
+\sum_{\alpha}b^{\dag}_{i\alpha}b_{i\alpha}\qquad \nonumber\\
& &+\sum_{\sigma\sigma^{\prime}}d^{\dag}_{i\sigma_{\alpha}\sigma_{\beta}}
 d_{i\sigma_{\alpha}\sigma_{\beta}}+q^{\dag}q
\end{eqnarray}
and
\begin{eqnarray}
\label{eq:numberconstrain}
\hat{Q}_{i\alpha\sigma}=f^{\dag}_{i\alpha\sigma}f_{i\alpha\sigma}.
\qquad\qquad\qquad\qquad\qquad\qquad\qquad
\end{eqnarray}

Our numerical method which is used to search the
minima energy is based on the pattern search method, the gradient
method and the Rosenbrock method. The normalization constraint given
by Eq.(\ref{eq:normalization}) must be held at all time in our
numerical calculation. The Fermion number constraint is enforced by
using the penalty function method. Our numerical calculations are
performed for simple square lattice with the nearest-neighbor
hopping. Throughout this paper we set the ratio of two intraorbital
hopping integrals as $t_{22}/t_{11}=0.5$, and the average electron
number per site is $2$. We measure the energies in units of the
bandwidth of the orbit-$1$,$2D_{1}=8t_{11}$.

\section{NUMERICAL RESULTS}

\begin{figure}[htbp]
\centering \setlength{\abovecaptionskip}{2pt}
\setlength{\belowcaptionskip}{4pt}
\includegraphics[angle=0, width=0.80 \columnwidth]{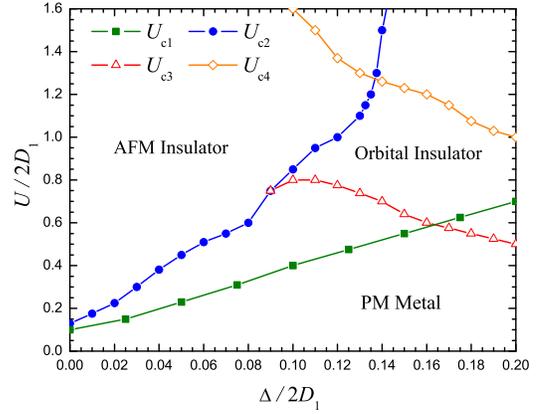}
\caption{Magnetic phase diagram in the $U-\Delta$
plane with $t_{12}=0$, where $U_{c1}$ and $U_{c2}$ are the PM metal
- AFM insulator boundary at $J_{H}=0.2U$ and $J_{H}=0.02U$
respectively. $U_{c3}$ and $U_{c4}$ are the orbital insulating phase
boundaries at $J_{H}=0.02U$ with and without spin degree of freedom
respectively.}
\label{fig:phasediagram}
\end{figure}

We first present the magnetic phase diagram of two-orbital Hubbard
model at half filling as the functions of $U$ and $\Delta$ in Fig.
1. For a large Hund's coupling, e.g. $J_{H}=0.2U$, there are only
paramagnetic (PM) metallic phase when $U<U_{c1}$ and AFM insulating
phase for $U>U_{c1}$ in the $U-\Delta$ phase diagram.
%
% Distinctly different from other PM results by DMFT and slave boson
% methods, the orbital selective Mott phase (OSMP) is not observed.
%
We notice that the magnetic MIT phase boundary $U_{c1}$ shifts
downward with the increase of the  crystal field splitting. When the
system transits from a PM metal to an AFM insulator, a fraction of
electrons transfer from a lower occupied orbit to an upper
unoccupied orbit. Consequently, the spin exchange splitting should
overcome the crystal field splitting, leading to the decrease in the
groundstate energy. We expect that the magnetic MIT is first order,
since it occurs only when the spin exchange splitting is superior to
the crystal field splitting. The orbital polarization induced by the
crystal field splitting is also expected to decrease sharply in the
AFM insulating phase.

For the small Hund's rule coupling situation with $J_{H}=0.02U$, the
magnetic phase diagram is richer than that with $J_{H}=0.2U$.
Besides the PM metallic and AFM insulating phases, an orbital
insulating state is stable when $U>U_{c3}$, {\it i.e.} the system
becomes completely orbital polarized without magnetism in the large
$\Delta$ region. As shown in Fig.\ref{fig:phasediagram}, the present
PM-AFM phase boundary $U_{c2}$ shifts to large $U$, in comparison
with the phase boundary $U_{c1}$ with $J_{H}=0.2U$. In the
comparable large $U$ and $\Delta$ and small $J_{H}$ region, two
electrons fall into the lower energy orbit. As a result, the system
undergoes an AFM/PM to orbital insulator transition.

The orbital insulating phase boundary without spin correlation is
also plotted as $U_{c4}$ in Fig.\ref{fig:phasediagram}, which is in
agreement with Werner and Millis' result \cite{Philipp2007}.
Obviously, the present critical value for orbital insulator,
$U_{c3}$, is considerably lower than that without spin correlation,
$U_{c4}$, though there is no sublattice magnetic moment in the
orbital insulating phase. The reason is that the modulation of spin
correlation on electronic spectrum is in favor of the occurrence of
the orbital insulator.

\begin{figure}[htbp]
\centering
\setlength{\abovecaptionskip}{2pt}
\setlength{\belowcaptionskip}{4pt}
\includegraphics[angle=0, width=0.85 \columnwidth]{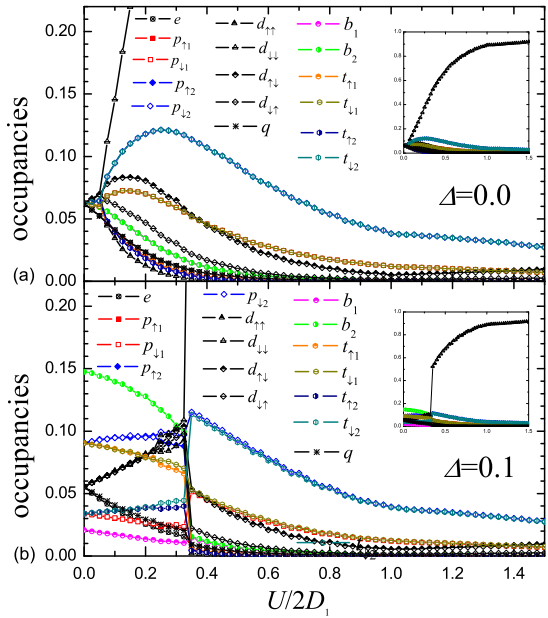}
\caption{Boson occupancy probabilities of sublattice as functions of $U$
for different crystal field splitting at $n=2$. Theoretical parameters:
$J_{H}=0.2U$, $t_{12}=0.2t_{11}$. (a) $\Delta$=0; (b) $\Delta$=0.1.
Insets in (a) and (b) show the full scale of the dependence
of occupancies on $U$}
\label{fig:occupancy}
\end{figure}

 To understand the electron distributions more clearly, we present
the dependence of the boson occupancy probabilities as the functions
of $U$ for different crystal field splittings in
Fig.\ref{fig:occupancy}. In the case of $J_{H}=0.02U$, the
spin-singlet double occupation probability $b_{2}$ is dominant over
the whole correlation range when the crystal field splitting
$\Delta$ is comparably large. In the case of $J_{H}=0.2U$, the
spin-triplet double occupation probability $d_{\sigma\sigma}$ is
dominant, as seen in Fig.\ref{fig:occupancy}.
Fig.\ref{fig:occupancy}(a) shows that when the orbital splitting
$\Delta=0$, the spin-up and spin-down occupancies split and the
ground state of the system is AFM at finite $U$. From the inset of
Fig.\ref{fig:occupancy}(a), one notices that
$d_{\downarrow\downarrow}$ considerably rises with the increase of
$U$, since the large $U$ and $J_{H}$ are in favor of the electron
distribution on different orbits with parallel spin, resulting in
the dominantly increase of $d_{\downarrow\downarrow}$. And the
evolution of various orbital occupancies probabilities on electron
correlation qualitatively agrees with Hasegawa's results
\cite{Hasegawa1997}.

However, in the presence of crystal field splitting, the boson
occupancy probabilities shown in Fig.\ref{fig:occupancy}(b) are very
different from Fig.\ref{fig:occupancy}(a). With increasing $U$, the
system undergoes a first-order MIT transition and transits from PM
to AFM phase at $U=U_{c1}$ since spin exchange splitting becomes
larger than crystal field splitting and spin gap comes into being.
At the same time, the spin-up and spin-down occupancies do not split
until $U>U_{c1}$. When $U>U_{c1}$, various occupancies monotonically
decrease with increasing $U$, except for $d_{\downarrow\downarrow}$,
which steeply increases and gradually approaches a saturation, as
one can see in the inset of Fig.\ref{fig:occupancy}(b).

\begin{figure}[htbp]
\centering
\setlength{\abovecaptionskip}{2pt}
\setlength{\belowcaptionskip}{4pt}
\includegraphics[angle=0, width=0.49 \columnwidth]{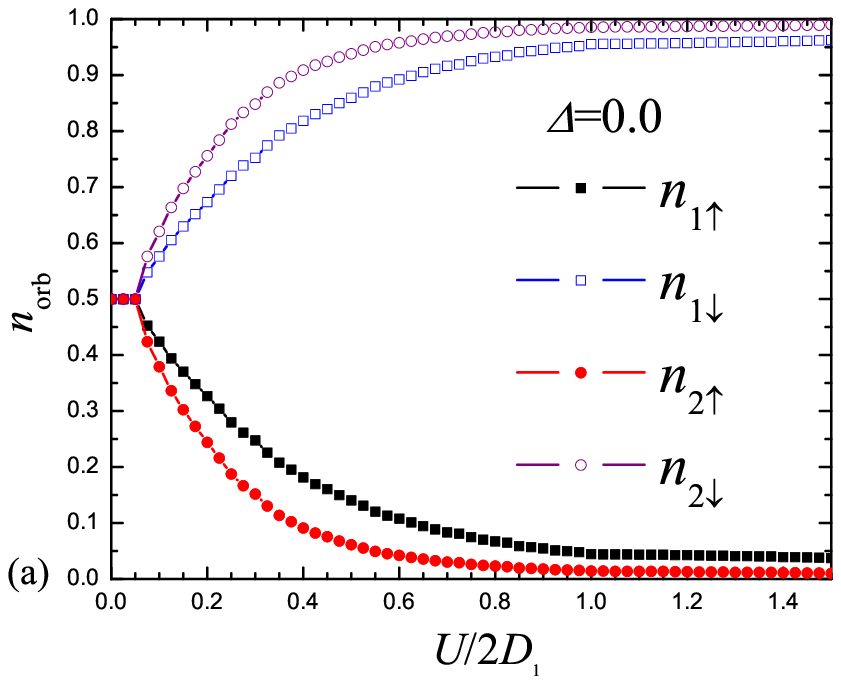}
\includegraphics[angle=0, width=0.49 \columnwidth]{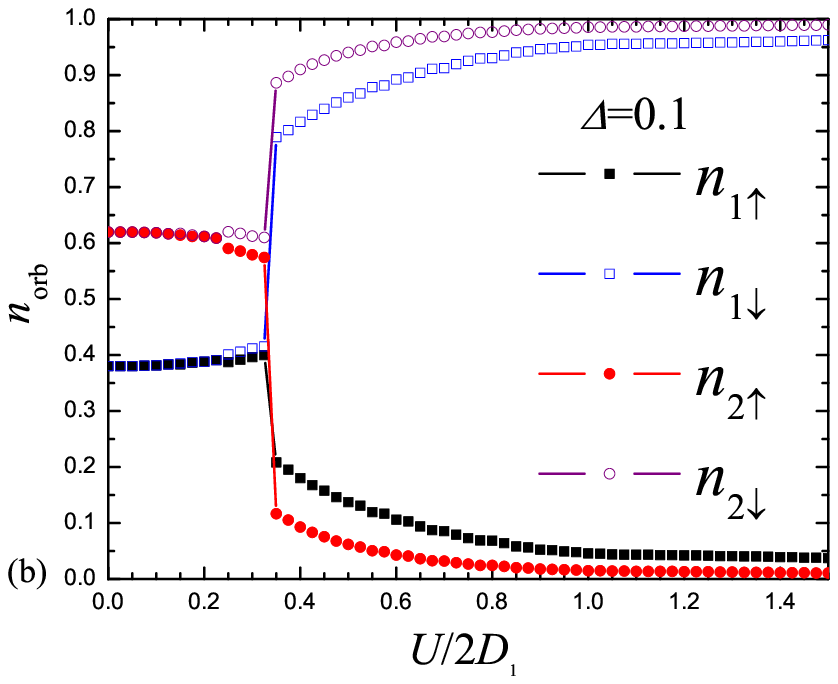}
\caption{Evolution of Electron number distributions in  two orbits.
(a) $\Delta=0$ and (b) $\Delta=0.1$. Other
parameters are the same to Fig.\ref{fig:occupancy}}
\label{fig:distributions}
\end{figure}

The evolution of the particle distributions in two orbits is plotted
in Fig.\ref{fig:distributions}. Fig.\ref{fig:distributions}(a) shows
that the sublattice magnetic moment contributed from narrow band is
lager than that from the wide band, since the spin exchange
splitting is stronger in the narrow band. In the present two-orbital
system with $\Delta=0$ and large $J_{H}$ in
Fig.\ref{fig:distributions}(a), the orbital polarization is always
suppressed since the gravity centers of two bands do not shift with
each other, the orbital singlet occupation dominates both in the PM
metallic and in the AFM insulating phases. And from
Fig.\ref{fig:distributions}b, it is found that although the crystal
field splitting induces the partial orbital polarization in the PM
metallic phase, it leads to completely unpolarized in the AFM
insulating phase since the each localized electron is allocated each
orbits, eliminating the orbital polarization.
%
% In addition, the magnetic moment contributions of two orbits are
% also different with crystal field.
%
\begin{figure}[htbp]
\centering
\setlength{\abovecaptionskip}{2pt}
\setlength{\belowcaptionskip}{4pt}
\includegraphics[angle=0, width=0.49 \columnwidth]{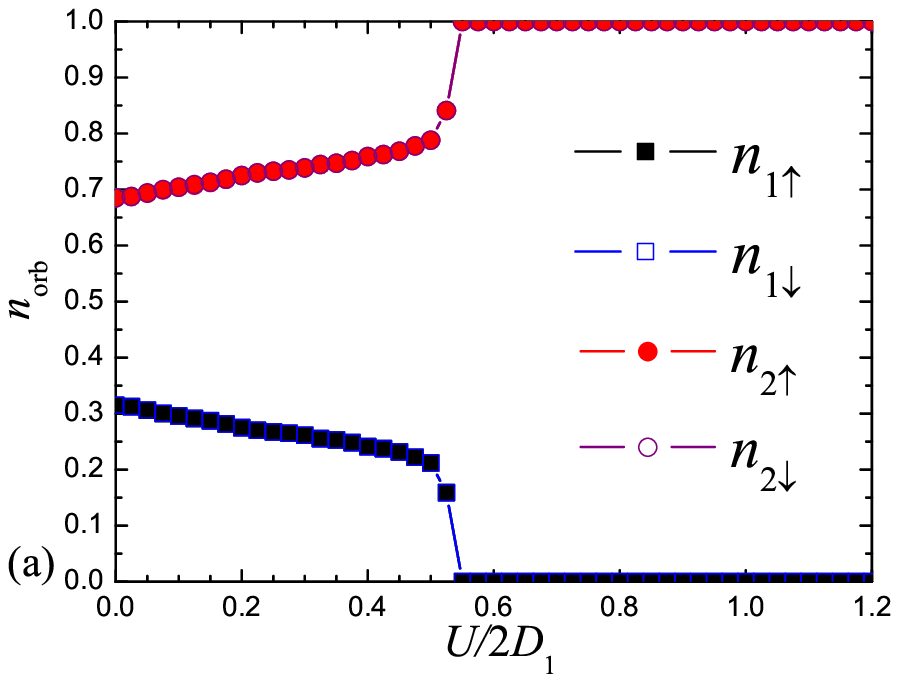}
\includegraphics[angle=0, width=0.49 \columnwidth]{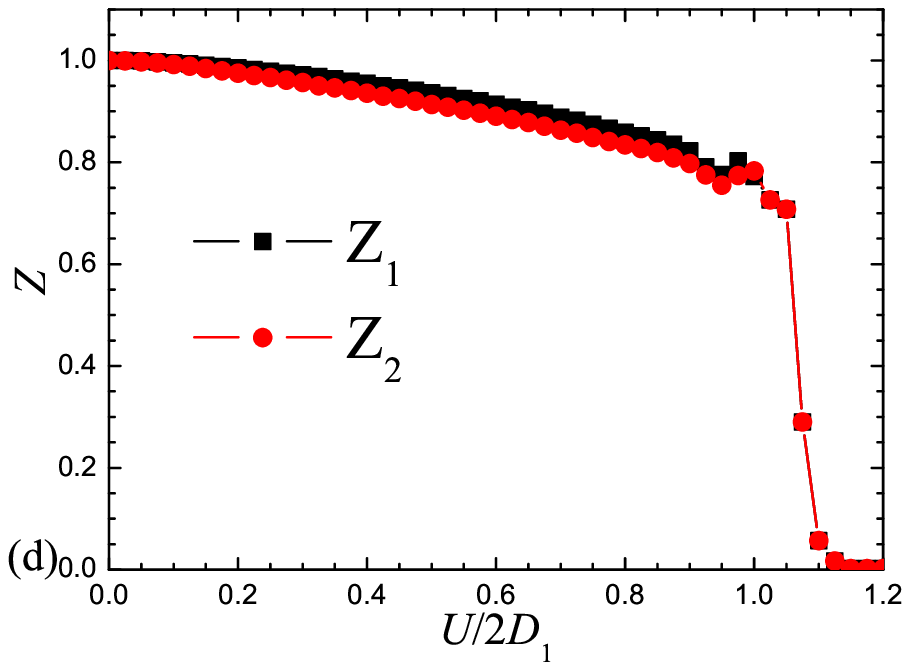}
\includegraphics[angle=0, width=0.49 \columnwidth]{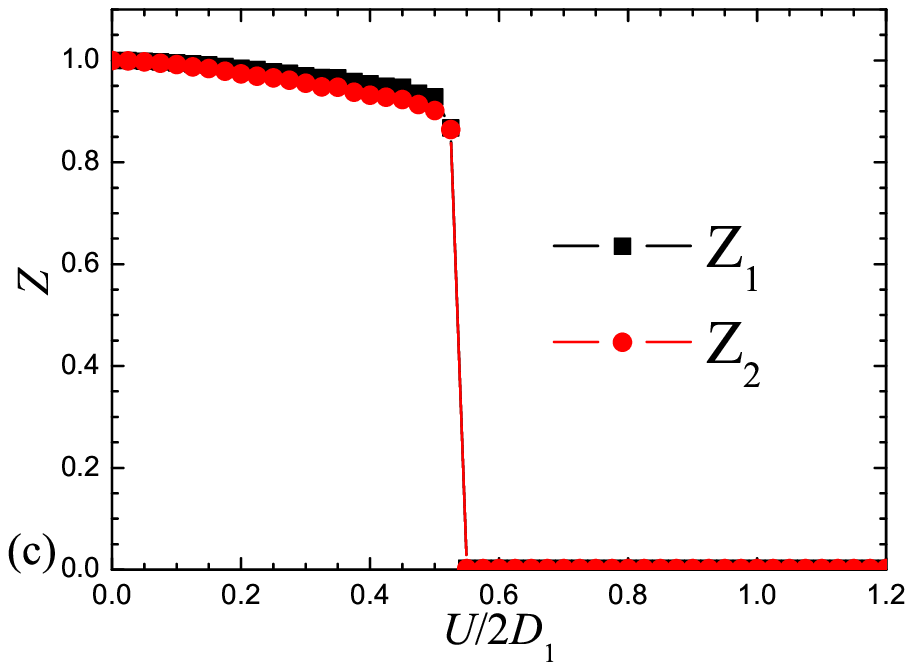}
\includegraphics[angle=0, width=0.49 \columnwidth]{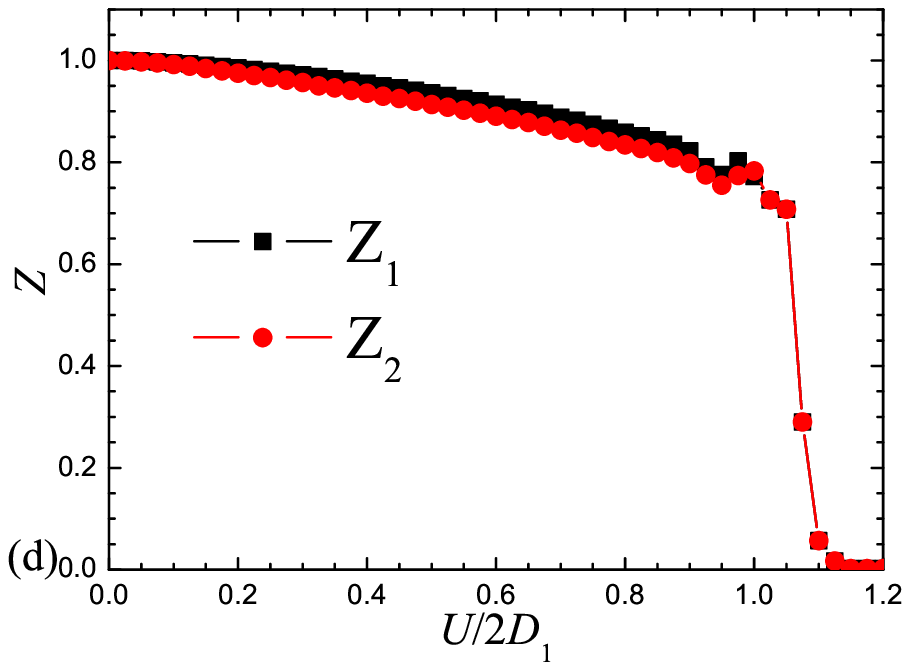}
\caption{Dependence of electron occupation numbers and
band renormalization factors of two orbits on Coulomb
correlation. (a) and (c) with spin correlation,
(b) and (d) without spin correlation. Theoretical
parameters: $\Delta=0.18$, $J_{H}=0.02U$, $t_{12}=0$.}
\label{fig:smalljh}
\end{figure}

The correlation dependence of the particle distributions and the
renormalizations factors of two orbits at $J_{H}=0.02U$ and
$\Delta=0.18$ are plotted in Fig.\ref{fig:smalljh}(a)and (c). The
corresponding results without considering spin exchange splitting
are also plotted in Fig.\ref{fig:smalljh}(b)and (d) for comparison.
Both Fig.\ref{fig:smalljh}(a) and \ref{fig:smalljh}(b) show that the
difference of particle numbers in two orbits, i.e. orbital
polarization, increases with the increase of $U$, and the system
becomes completely orbital polarized in the orbital insulating phase
when $U>U_{c3}$, accompanied with the PM-AFM MIT. The corresponding
renormalization factors in Fig.\ref{fig:smalljh}(c) and
\ref{fig:smalljh}(d) show that the MIT simultaneously occur in the
two orbits, even if in the presence of finite crystal field
splitting. Comparing Fig.\ref{fig:smalljh}(a) with
Fig.\ref{fig:smalljh}(b) and Fig.\ref{fig:smalljh}(c) with
Fig.\ref{fig:smalljh}(d), we find that the spin correlation greatly
reduces the critical value $U_{c}$ of the MIT, therefore favors the
occurrence of the MIT.

Our numerical results for different inter-orbital hopping do not
qualitatively affect the magnetic MIT, only the phase boundaries
show a slight shift. We find that different from the statement by Yu
and Si \cite{Qimiao2010}, in addition to $J_{H}=0$, the orbital
insulating phase can exist for finite but small $J_{H}$, since the
competition between crystal field splitting and the Hund's rule
coupling may stabilize the orbital insulating phase. It is
interesting to expect that considerable lattice distortion may be
associated with the occurrence of orbital insulator due to the
first-order phase transition with large variation in orbital
polarization. Further, as seen in Fig.\ref{fig:phasediagram},
\ref{fig:occupancy} and \ref{fig:distributions}, the high-spin to
low-spin state transition is clearly found in our present study,
proving Millis \emph{et al}.'s conjecture in their paramagnetic DMFT
study \cite{Philipp2007}, which indicates that the MIT with
magnetism is much more rich than that without magnetism.

\section{SUMMARY AND CONCLUSION}

We have performed a comparative study on the alternative effects of
crystal field splitting and Hund's rule coupling on MIT of
half-filled two-orbital asymmetric Hubbard model with and without
magnetism by means of the Kotliar-Ruckenstein slave boson approach.
Our results show that the MIT is first order accompanied with a
magnetic transition in the presence of crystal field splitting. The
critical value $U_{c}$ for MIT considerably reduces after taking into
account the spin correlation. Meanwhile, the orbital insulator can
exist in the system with large crystal field splitting and small
Hund's rule coupling. With the increase of crystal field splitting,
the system may enter from a high-spin AFM state to a low-spin
orbital insulating phase in the small $J_{H}$ region. These results
demonstrate that the competition between the crystal field splitting
and the Hund's rule coupling $J_{H}$ is very important for the
metal-insulator transition with spin degree of freedom.

\acknowledgements

This work was supported by the National Science Foundation of China
under Grant No. 10874186 and 11074257, Knowledge Innovation Program
of Chinese Academy of Sciences, and Director Grants of CASHIPS.
Numerical calculations were performed in Center for Computational
Science of CASHIPS.


\end{document}